\documentclass{article}
\usepackage{spconf,amsmath,graphicx,hyperref}
\usepackage{amssymb}
\usepackage{booktabs} 
\usepackage{multirow} 
\usepackage{subcaption}

\title{TLDiffGAN: A Latent Diffusion-GAN Framework with Temporal Information Fusion for Anomalous Sound Detection}
\name{Chengyuan Ma$^{1}$, Peng Jia$^{2}$, Hongyue Guo$^{2}$, Wenming Yang$^{1 \star}$
\thanks{$^{\star}$ Corresponding authors} 
\thanks{This work was supported in part by the National Key R\&D Program of China (2023YFB4302200) and the Special Foundations for the Development of Strategic Emerging Industries of Shenzhen (KJZD20231023094700001).}}
\address{$^1$Shenzhen International Graduate School, Tsinghua University, China\\ 
$^2$Collaborative Innovation Center for Transport Studies, Dalian Maritime University, China} 
\begin{document}
\ninept
\maketitle
\begin{abstract}

Existing generative models for unsupervised anomalous sound detection are limited by their inability to fully capture the complex feature distribution of normal sounds, while the potential of powerful diffusion models in this domain remains largely unexplored. To address this challenge, we propose a novel framework, TLDiffGAN, which consists of two complementary branches. One branch incorporates a latent diffusion model into the GAN generator for adversarial training, thereby making the discriminator’s task more challenging and improving the quality of generated samples. The other branch leverages pretrained audio model encoders to extract features directly from raw audio waveforms for auxiliary discrimination. This framework effectively captures feature representations of normal sounds from both raw audio and Mel spectrograms. Moreover, we introduce a TMixup spectrogram augmentation technique to enhance sensitivity to subtle and localized temporal patterns that are often overlooked. Extensive experiments on the DCASE 2020 Challenge Task 2 dataset demonstrate the superior detection performance of TLDiffGAN, as well as its strong capability in anomalous time–frequency localization.
\end{abstract}

\vspace{-1mm}
\section{Introduction}
\label{sec:intro}




Anomalous Sound Detection (ASD) aims to identify previously unseen anomalous events using only normal sound data \cite{DCASE2020_01}, and is of critical importance in applications such as intelligent monitoring of industrial equipment. Existing mainstream approaches can be broadly divided into two categories. The first category consists of classification-based self-supervised models \cite{Giri2020a,SW23,guan2023anomalous}, which can achieve strong performance under specific conditions but typically rely heavily on metadata such as machine IDs. Moreover, their decision-making processes lack inherent interpretability and are unable to localize anomalies in the time–frequency domain, which severely limits their generalizability. In light of these limitations, this study focuses on the second category—reconstruction-based generative models—that do not require metadata and inherently provide the potential for anomaly localization and interpretability through the analysis of reconstruction residuals.

Generative models identify anomalies by learning the distribution of normal sounds. However, early autoencoders (AEs) and their variants \cite{Suefusa_Nishida_Purohit_Tanabe_Endo_Kawaguchi_2020,kapka2020id, ANP} exhibit inherent limitations in their learning mechanisms, making it difficult to capture a  complete distribution of normal sound features. As a result, the models often learn a relatively broad reconstruction mapping that not only applies to normal sounds but can also partially accommodate structurally similar anomalous sounds, leading to insufficient reconstruction errors for anomalies. Subsequently, Generative Adversarial Networks (GANs) \cite{AnoGAN19,GANomaly24,aeganad23} were introduced, but their inherent training instability and risk of mode collapse limit their applicability in complex acoustic scenarios. More recently, Denoising Diffusion Probabilistic Models (DDPMs) \cite{ho2020denoising} have demonstrated remarkable generative capacity, but they also introduce new challenges. Their strong distribution learning ability may cause the model to interpret anomalous features as noise and remove them during reconstruction, thereby generating a distribution that differs only marginally from the anomalous input, which complicates anomaly detection. In addition, a frequently overlooked limitation is that nearly all of the above generative models rely exclusively on Mel spectrograms as input. However, spectrograms inevitably discard part of the critical information contained in raw waveforms during the time–frequency transformation \cite{ST22}. This drawback of single-modality input fundamentally constrains the upper bound of model performance. Furthermore, existing generative models tend to capture global and macroscopic characteristics of normal sounds during training. Such a learning paradigm may render the models insufficiently sensitive to weak transient variations that occur only within localized time–frequency regions.






To address the above challenges, this paper proposes a novel anomalous sound detection framework, aiming to overcome the inherent trade-offs among reconstruction fidelity, training stability, and anomaly sensitivity in existing generative models. By effectively integrating complementary acoustic information from both spectrograms and raw waveforms, the framework substantially enhances the anomaly detection capability of generative models and ultimately achieves superior time–frequency localization of anomalies. The main contributions of this work can be summarized as follows:

(1) We propose a dual-branch acoustic modeling framework that combines spectrogram reconstruction with raw waveform feature extraction. Through an innovative Latent Diffusion–GAN (LDGAN) and pretrained encoders, the framework effectively fuses complementary acoustic information from log-Mel spectrograms and raw waveforms.

(2) We design an adaptive temporal mixup (TMixup) module that leverages an attention mechanism to identify and enhance feature regions located at the boundary of the normal data distribution, thereby improving the model’s ability to discriminate along the decision boundary between normal and anomalous sounds.

(3) Extensive experiments on the DCASE 2020 Task 2 benchmark dataset demonstrate that our proposed framework significantly outperforms mainstream generative models across multiple key metrics, validating both the effectiveness of its design and its superior performance.

\vspace{-2mm}
\section{Proposed Method}
\begin{figure}[htb]
  \centering
  \includegraphics[width=0.5\textwidth]{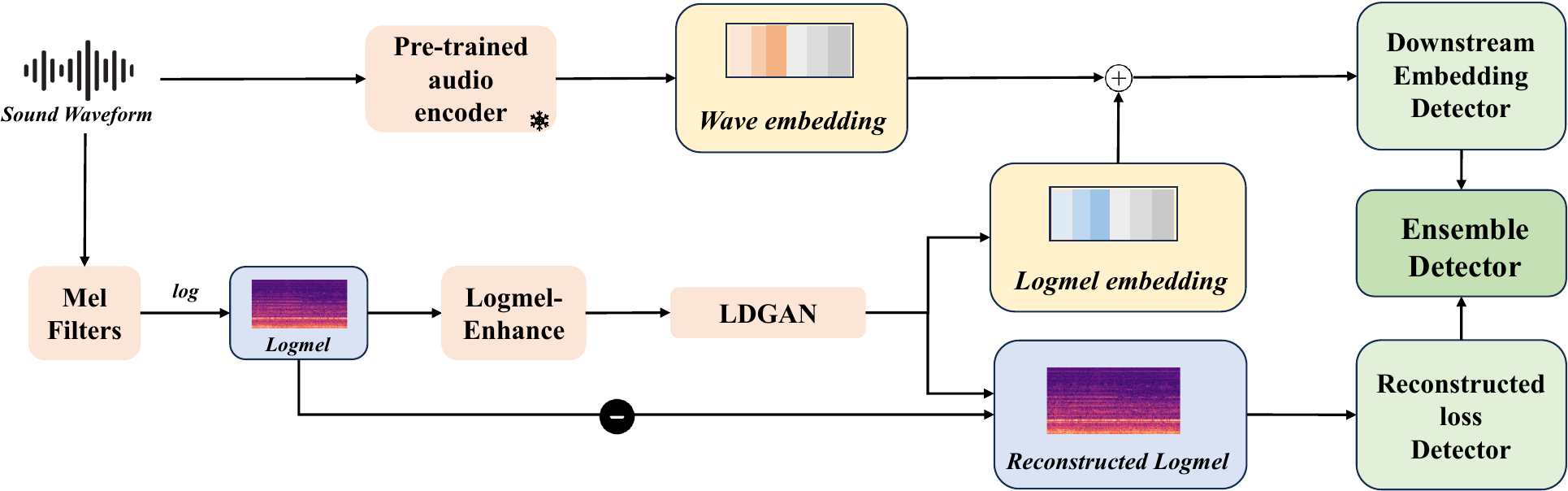}
  \makeatletter
  \def\@makecaption#1#2{%
    \vskip\abovecaptionskip
    \centering
    {\small \normalfont #1.~#2\par}
    \vskip\belowcaptionskip}
  \makeatother
    \vspace{-2mm}
  \caption{The main pipeline of our TLDiffGAN.}
  \vspace{-2mm}
  \label{fig:1}
\end{figure}

\vspace{-2mm}
The pipeline of our proposed method is illustrated in Fig. \ref{fig:1}. This section provides a comprehensive explanation of the details. The Latent Diffusion-GAN backbone module for log-mel spectrogram feature extraction is described in Section 2.1. The TMixup module, which is designed to enhance the temporal features of log-mel spectrograms, is discussed in Section 2.2. The pretrained audio encoder module utilized in our approach is outlined in Section 2.3, and finally, the detector component of our method is introduced in Section 2.4.

\vspace{-2mm}
\subsection{LDGAN Backbone}
  \vspace{-2mm}

\begin{figure}[ht] 
\centering
\includegraphics[width=0.9\linewidth]{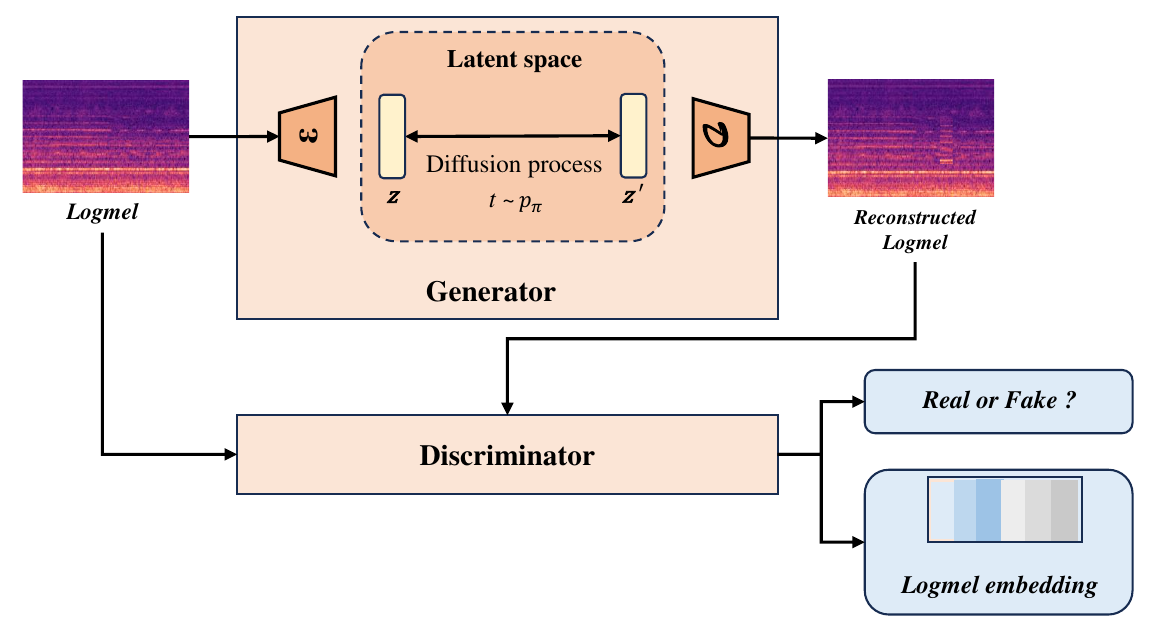}
  \makeatletter
  \def\@makecaption#1#2{%
    \vskip\abovecaptionskip
    \centering
    {\small \normalfont #1.~#2\par}
    \vskip\belowcaptionskip}
  \makeatother
  \vspace{-2mm}
\caption{LDGAN Framework.}
\vspace{-2mm}
\label{fig:2}
\end{figure}

At the core of our framework lies an innovative LDGAN backbone (Fig. \ref{fig:2}), which deeply integrates the stepwise denoising capability of Latent Diffusion Models (LDM)  with the adversarial training mechanism of Generative Adversarial Networks (GANs) to reconstruct high-quality Mel spectrograms of normal sounds. Unlike conventional GAN generators that directly model the data distribution, our generator learns to synthesize spectrograms through a progressive reverse denoising process from noise to data. This process is carried out in a low-dimensional latent space and guided by the GAN discriminator, with the design aiming to integrate the high generative quality of diffusion models with the training stability of the GAN framework.

During training, the generator is optimized with a dual objective: it must both accurately learn the reverse denoising trajectory of the diffusion model and ensure that the final generated samples are indistinguishable from real samples in the feature space. Accordingly, the overall generator loss $\mathcal{L}_G$ is defined as a weighted combination of the noise prediction loss $\mathcal{L}_{\mathrm{noise}}$ and the statistical matching loss $\mathcal{L}_{\mathrm{stat}}$:
\begin{equation}
\mathcal{L}_G=\mathcal{L}_\mathrm{noise}+\lambda_\mathrm{stat}\cdot\mathcal{L}_\mathrm{stat}
\end{equation}
Specifically, the noise prediction loss $\mathcal{L}_{\mathrm{noise}}$ follows the standard LDM objective, which minimizes the $L_2$ error between the predicted noise and the ground-truth noise. In contrast, the statistical matching loss $\mathcal{L}_{\mathrm{stat}}$ aligns the feature distributions of real and generated samples by minimizing the $L_2$ distance between their deep features extracted by the discriminator.

The discriminator $D$, in turn, faces a more challenging task: it must not only distinguish between real and final generated spectrograms but also evaluate intermediate denoising steps, thereby providing gradient signals for the generator’s refinement at each stage. To stabilize training, we apply spectral normalization to $D$ and incorporate a gradient penalty term $\mathcal{L}_{\mathrm{GP}}$ \cite{wgangp}. The discriminator’s total loss is formulated as:
\begin{equation}
\mathcal{L}_D=\mathcal{L}_{\mathrm{adv}}+\lambda_{\mathrm{GP}}\cdot\mathcal{L}_{\mathrm{GP}}
\end{equation}
where $\mathcal{L}_{\mathrm{adv}}$ denotes the standard adversarial loss.

By integrating the latent diffusion process into the GAN framework, the generator learns reconstruction along a structured, progressive refinement path, while the adversarial objective drives the synthesis of perceptually sharper and more realistic spectral details. This effectively overcomes the reconstruction blurriness commonly observed in autoencoder-based approaches. Moreover, the structured nature of diffusion provides a strong regularization effect for adversarial training, thereby mitigating the inherent risk of mode collapse in GANs.

\vspace{-2mm}
\subsection{Log-mel Spectrogram Enhancement}


To enhance the temporal feature representation in log-mel spectrograms, we improved the module from \cite{Noisy24} by introducing power average pooling and trainable weights. Specifically, we define three pooling operations: Max Pooling, Average Pooling, and Power Average Pooling \cite{liu2021power}, and perform a weighted summation of their results using trainable weights $\hat{w}_{\mathrm{max}}$, $\hat{w}_{\mathrm{avg}}$, and $\hat{w}_{\mathrm{pow}}$.
\begin{equation}
\begin{aligned}
&\mathrm{Pool}(x_{\mathrm{mel}})=\hat{w}_{\mathrm{max}}\cdot\mathrm{Max}\mathrm{Pool}(x^T_{\mathrm{mel}}) 
+\hat{w}_{\mathrm{avg}}\cdot\mathrm{Avg}\mathrm{Pool}(x^T_{\mathrm{mel}})\\
&+\hat{w}_{\mathrm{pow}}\cdot \mathrm{Power}\text{AvgPool}(x^T_{\text{mel}})
\end{aligned}
\end{equation}
where $x_{\mathrm{mel}} \in \mathbb{R}^{F \times T}$ represents the log-mel spectrogram, where $F$ and $T$ denote the frequency and time dimensions, respectively; $\hat{w}_{\mathrm{max}}$, $\hat{w}_{\mathrm{avg}}$, and $\hat{w}_{\mathrm{pow}}$ are trainable weight parameters, which are normalized using softmax to ensure their sum equals 1. These weights are automatically adjusted during the training process through backpropagation to optimize the extraction of temporal features.  To further optimize the pooling results, we first apply a Sigmoid activation function to transform the weighted pooling output into a soft temporal attention map $x_{TA} \in \mathbb{R}^{T}$, whose values lie within $(0,1)$ and quantify the confidence of each time frame deviating from the normal pattern. Subsequently, to accurately localize and select the most suspicious regions for enhancement, we convert this soft attention map into a hard mask $M_{\mathrm{mask}} = \mathbb{I}(x_{TA} > \tau)$ using a threshold $\tau$ that is randomly sampled during each training step, where $\mathbb{I}(\cdot)$ denotes the indicator function that explicitly specifies the regions for the subsequent local Mixup operation. Subsequently, we perform Mixup localized mixing operations on these high-attention regions:
\begin{equation}
    x_{\mathrm{mixed}}=\lambda\cdot x_{\mathrm{mel}}+(1-\lambda)\cdot\left[x_{\mathrm{mel}}\otimes M_{\mathrm{mask}}\right]
\end{equation}
where the mixing coefficient $\lambda\sim\mathrm{Beta}(\alpha,\alpha)$  controls the weighted mixing ratio between the original spectrogram and the localized temporal attention features, and $\otimes$ represents element-wise multiplication. By using log-mel spectrograms in the TMixup module, the model is able to capture features across multiple time scales, effectively detecting regions that deviate from the normal pattern. We deliberately amplify the model’s sensitivity to subtle feature variations by applying local augmentation to regions on the boundary of the normal data distribution. This encourages the model to learn to focus on the boundaries of the normal range during training, thereby enabling more effective detection of true anomalies at inference.

\vspace{-2mm}
\subsection{Pre-trained Audio Encoder}
\vspace{-2mm}


To compensate for the potential information loss in Mel spectrograms, we design a parallel branch that leverages pretrained self-supervised learning (SSL) models to extract deep features directly from raw audio waveforms. The acoustic characteristics of industrial equipment sounds often differ substantially from speech signals, typically exhibiting long-term stationary patterns and sparse transient events. Hence, general-purpose audio pretrained models are better suited for this task than speech-specific models.

In this work, we mainly investigate powerful Transformer-based architectures. For example, AST \cite{ast} and its self-supervised variant ATST \cite{atst} adapt the Vision Transformer architecture to audio spectrograms, enabling efficient modeling of the global contextual information of sound events. Another representative model, BEATs \cite{beats}, employs an iterative pretraining framework with a learned audio tokenizer to distill semantically richer, high-level acoustic representations. In addition, efficient audio Transformers such as EAT \cite{eat} focus on capturing long-range dependencies, which are crucial for analyzing subtle acoustic variations between stable operating states and anomalous conditions. By integrating these encoders, our framework obtains robust embeddings from raw waveforms, thereby preserving critical acoustic details that may otherwise be smoothed out or overlooked during the conversion to Mel spectrograms.

\vspace{-3mm}
\subsection{Detector}


Our detection framework consists of two parallel components: a reconstruction-based detector and an embedding-based detector. First, the reconstruction-based detector aims to quantify how well a given sample conforms to the latent manifold of normal sounds learned by the model. When the input is anomalous, its features deviate from the learned normal distribution, making accurate reconstruction difficult and thus yielding large discrepancies. The anomaly score $s_{r}$ is computed as the squared Euclidean distance between the latent representations of the real and reconstructed samples:

\begin{equation}
s_r = ||z_{\mathrm{real}} - z_{\mathrm{rec}}||_2^2
\end{equation}

Second, the embedding-based detector operates in a high-dimensional joint feature space ($\mathcal{Z}$), formed by concatenating Mel-spectrogram features $\mathcal{Z}_{\mathrm{mel}}$ with raw waveform features $\mathcal{Z}_{\mathrm{wave}}$. To comprehensively capture different types of anomalous patterns, we adopt a set of complementary classical algorithms: K-Nearest Neighbors (KNN) \cite{knn} for measuring the distance of a sample to the normal feature manifold; Local Outlier Factor (LOF) \cite{lof} for identifying samples with abnormal local density; Gaussian Mixture Model (GMM) \cite{GMM} for estimating the likelihood of a sample under the complex distribution of normal sounds; and Subspace Outlier Score (SOS) \cite{sos} for detecting subtle anomalies that become salient only within specific feature subspaces. Our method employs a validation-based metric selection strategy. Specifically, for each machine type, we independently select the best-performing detector based solely on validation performance; this configuration remains fixed during inference, ensuring the process operates without reliance on machine IDs. The final anomaly score is determined as follows:

\begin{equation}
s_{\mathrm{final}}^{(k)}=
\begin{cases}
s_r & \mathrm{if~}\mathcal{M}_k^{(r)}=\max(\{\mathcal{M}_k^{(m)}\}) \\
s_{e_1} & \mathrm{if~}\mathcal{M}_k^{(e_1)}=\max(\{\mathcal{M}_k^{(m)}\}) \\
\vdots & \vdots \\
s_{e_4} & \mathrm{if~}\mathcal{M}_k^{(e_4)}=\max(\{\mathcal{M}_k^{(m)}\})  
\end{cases}
\end{equation}

where $\mathcal{M}_k^{(m)}$ denotes the overall performance metric of the $m$-th detector on the $k$-th machine type, defined as the arithmetic mean of AUC and pAUC.

\vspace{-2mm}

\section{Experiment and Results}
\label{sec:EXPERIMENT and result}

\subsection{Experimental Setup}
\noindent \textbf{Dataset:} We conduct experiments on the DCASE 2020 Challenge Task 2 dataset \cite{MIMII_DCASE2019_01, ToyADMOS2019_01}, which consists of two subsets, MIMII and ToyADMOS, covering six machine types: Fan, Pump, Slider, Valve, ToyCar, and ToyConveyor. Each audio sample is monaural, recorded at a sampling rate of 16 kHz, with a duration of approximately 10 seconds. We did not adopt later versions of the DCASE datasets, as they primarily emphasize challenges related to domain shift. In contrast, the core focus of this work is to model the distribution of normal sounds within a single domain.


\noindent \textbf{Evaluation Metrics:} We adopt the Area Under the ROC Curve (AUC) and the partial AUC (pAUC) under a controlled false positive rate as evaluation metrics. Following the official challenge protocol \cite{DCASE2020_01}, pAUC is computed within the low false positive rate interval $[0, p]$, where $p$ is set to 0.1. For all metrics, higher values indicate better performance.

\noindent \textbf{Implementation Details:} For the input features, the raw audio waveforms are converted into log-mel spectrograms with dimensions of $128 \times 313$. In the TMixup module, the threshold $\tau$ of the temporal attention map is randomly sampled from a uniform distribution $U(0.2, 0.5)$. Regarding network configuration, the discriminator in our LDGAN backbone shares a similar structure with the encoder but employs grouped convolutions in the final layer to extract deep semantic features. The loss weighting coefficients $\lambda_\mathrm{stat}$ and $\lambda_\mathrm{{GP}}$ were set to 1.0 and 10, respectively. The model is trained using the Adam optimizer with a learning rate of 0.0001, a batch size of 512, and a total of 150 epochs. A gradient penalty is applied to the discriminator to further enhance training stability.

\vspace{-2mm}
\subsection{Performance Comparison}

\begin{table*}[htb]
\label{table}
\caption{Performance Comparison with Other Methods on the DCASE 2020 Task 2 Dataset. Official baseline: the autoencoder baseline provided by DCASE. Average: the average of the AUC and pAUC values for all machine types.}
  \vspace{-2mm}
    \resizebox{\textwidth}{!}{
    \scalebox{0.9}{
    \begin{tabular}{lcccccccccccc|ccc}
        \toprule
        \multirow{2}{*}{Methods} & \multicolumn{2}{c}{Fan} & \multicolumn{2}{c}{Pump} & \multicolumn{2}{c}{Slider} & \multicolumn{2}{c}{ToyCar} & \multicolumn{2}{c}{ToyConveyor} &  \multicolumn{2}{c}{Valve} & \multicolumn{2}{c}{Average}\\
        \cmidrule(lr){2-3} \cmidrule(lr){4-5} \cmidrule(lr){6-7} \cmidrule(lr){8-9} \cmidrule(lr){10-11} \cmidrule(lr){12-13} \cmidrule(lr){14-15} 
         & AUC & pAUC & AUC & pAUC & AUC & pAUC & AUC & pAUC & AUC & pAUC  & AUC & pAUC & AUC & pAUC \\
        \midrule        
        Official baseline \cite{DCASE2020_01}    & 
        65.91 & 52.74 & 71.36 & 60.02 & 83.86 & 66.42 & 78.23 & 67.38 & 71.01 & 59.78 & 65.28 & 49.98 & 72.61 & 59.39\\

        ANP \cite{ANP}    & 
        69.20 & 54.40 & 72.80	&61.80&	90.70	&74.20	&86.90	&70.70	&72.50	&67.30	&67.00&	54.50&	76.52	&63.82\\


        GANomaly \cite{GANomaly24} & 
        79.37 & 63.48 & 72.65 & 61.48 & 84.21 & 72.84  & 85.12 & 72.23 & 74.59 & 61.24 & 79.30 & 57.74 & 79.21 & 64.84 \\

        ASD-Diffusion \cite{zhang2025asd} & 
        83.64 & 71.92 & 82.78 & 74.92 & 88.51 & 75.24  & 92.30 & 81.48 & 78.65 & 63.12 & 87.78& 61.55 & 85.61 & 71.37 \\
        
        AEGAN-AD \cite{aeganad23} & 
        83.12 & 71.86 & 84.37 & 75.42 & 91.84 & 78.18  & 91.70 & 80.40 & 79.00 & \textbf{65.86} & 84.37 & 60.75 & 86.08 & 72.08 \\

        Ours & 
        \textbf{85.88} & \textbf{73.15} & \textbf{87.60} & \textbf{76.55} & \textbf{94.78} & \textbf{83.94}  & \textbf{93.35} & \textbf{82.97} & \textbf{80.29} & 65.21 & \textbf{89.67} & \textbf{64.26} & \textbf{88.60} & \textbf{74.35} \\
     
        \bottomrule
        \bottomrule
    \end{tabular}
    }
    }
    \label{tab:1}
    \vspace{-2mm}
\end{table*}

\vspace{-1mm}

As shown in Table \ref{tab:1}, we compared the performance of TLDiffGAN with other competing methods. Since generative models do not rely on metadata and can localize anomalous regions, we compared the performance of TLDiffGAN with other competing generative methods, particularly on the Slider and Pump datasets, where AUC and pAUC were improved by 2.94\%, 5.76\%, and 3.23\%, 1.13\%, respectively. Although AEGAN-AD showed slightly better pAUC performance on the ToyConveyor dataset, our proposed method delivered superior detection performance in terms of AUC and pAUC across the vast majority of machine types. Additionally, TLDiffGAN achieved the best average AUC and pAUC scores of 88.60\% and 74.35\%, respectively, further proving the superiority of our model.

\vspace{-2mm}

\subsection{Comparison between Pre-trained Audio Encoder}

Different pretrained models exhibit varying levels of performance on ASD tasks due to differences in their architectural designs. As shown in Table \ref{tab:2}, experiments conducted by replacing only the pretrained audio encoder indicate that EAT achieves the best performance, followed by BEATs, while other models perform relatively poorly. EAT’s large-block unmasking strategy further enhances its ability to capture long-term dependencies, making it particularly suitable for detecting the subtle anomalous features often present in machine sounds.

\begin{table}[!ht]
\small
\vspace{-2mm}
\centering
\setlength{\tabcolsep}{3pt}
\caption{Average AUC (\%) and pAUC (\%) comparison of different pre-trained audio encoders across six machine types.}
\vspace{-2mm}
\scalebox{0.85}{
\begin{tabular}{c c cc}
    \toprule
    \multirow{2}{*}{Model} & \multirow{2}{*}{Size} & \multicolumn{2}{c}{Average} \\ 
    \cmidrule(lr){3-4}
     & & AUC & pAUC \\
    \midrule
    AST \cite{ast} & 86M & 85.47 & 71.40 \\
    
    ATST \cite{atst} & 85M & 85.85 & 71.24 \\
    BEATs \cite{beats}   & 90M & 86.92 & 73.98 \\
    EAT \cite{eat} & 88M & \textbf{88.60} & \textbf{74.35} \\
    \bottomrule
\end{tabular}
}
\label{tab:2}
\vspace{-2mm}
\end{table}

\vspace{-2mm}
\subsection{Ablation Study}

To evaluate the contribution of each core component within the framework, we conducted ablation studies, with the results summarized in Table \ref{tab:3}. The averaged performance results clearly indicate that removing any of the latent diffusion, the EAT encoder, or the log-mel spectrogram enhancement module leads to a degradation in overall model performance. For sound sources such as Fan, which exhibit highly stationary and simple acoustic patterns, incorporating additional raw waveform features may introduce redundant information. In contrast, for sources like ToyConveyor, whose normal operation inherently contains complex temporal patterns, the strong augmentation effect of the TMixup module may slightly perturb the learning of the normal distribution. Nevertheless, the complete proposed model consistently achieves the best overall average performance across all machine types, thereby validating the effectiveness of our approach.

\begin{table}[!ht]
\small
\vspace{-2mm}
\centering
\setlength{\tabcolsep}{3pt}

\caption{Ablation study on the core components of TLDiffGAN, with results in AUC (\%).}
\vspace{-2mm}
    \scalebox{0.8}{
    \begin{tabular}{ccccc}
        \toprule
         Model & TLDiffGAN & \begin{tabular}[c]{@{}c@{}}w/o\\ Latent Diffusion\end{tabular} & \begin{tabular}[c]{@{}c@{}}w/o\\ EAT\end{tabular} & \begin{tabular}[c]{@{}c@{}}w/o\\ Logmel-Enhance\end{tabular}   \\
         \midrule
         Fan & 85.88 & 83.27 & \textbf{86.35} & 85.14  \\
         Pump & \textbf{87.60} & 85.95 & 87.28 & 86.55\\
         Slider & \textbf{94.78} & 92.52 & 90.94 & 91.69 \\
         ToyCar & \textbf{93.35} & 92.14 & 88.51 & 93.10 \\
         ToyConveyor & 80.29 & 79.85 & 78.32 & \textbf{80.86} \\
         Valve & \textbf{89.67} & 84.68 & 86.73 & 88.50 \\
          \hline
         Average & \textbf{88.60} & 86.40 & 86.36 & 87.64 \\
         \bottomrule
    \end{tabular}
    }
    \label{tab:3}
    \vspace{-2mm}
\end{table}
\vspace{-2mm}

\subsection{Localization of Spectrogram Regions}
\begin{figure}[h]
  \centering
  \begin{minipage}{0.8\linewidth} 
    \centering
    \includegraphics[width=\linewidth]{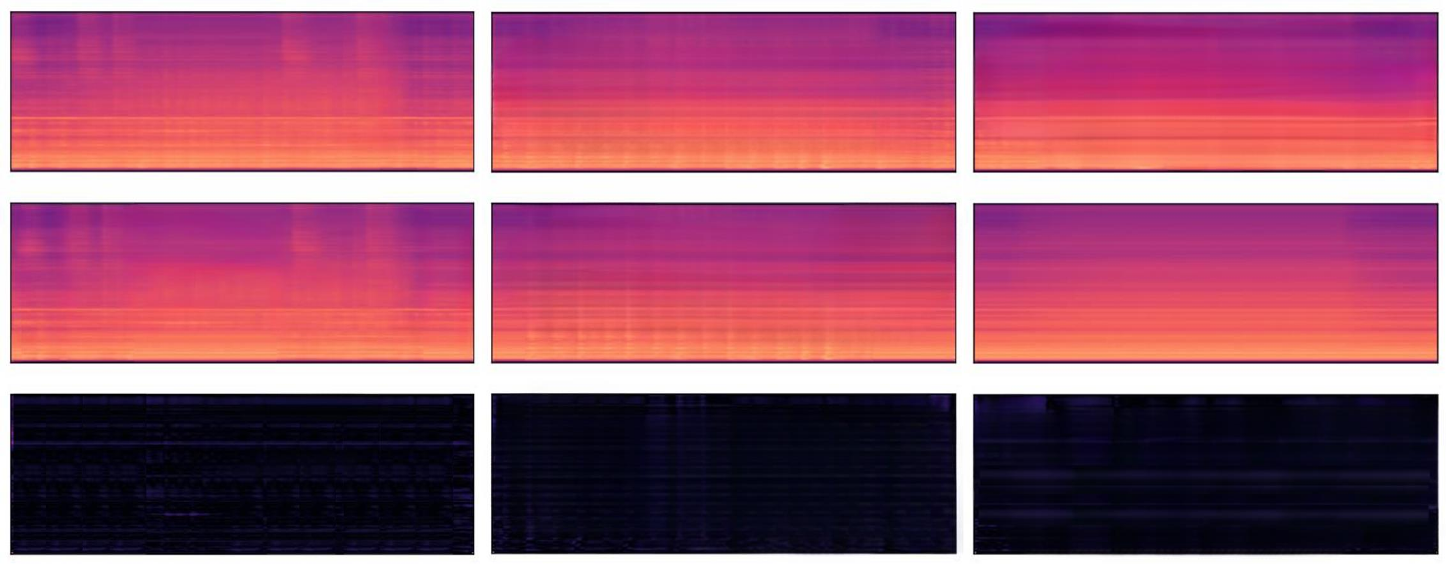}
    {\small (a) ToyCar Normal }
    \label{fig:sub1}
  \end{minipage}
  \vspace{-0.5mm}
  
  \begin{minipage}{0.8\linewidth} 
    \centering
    \includegraphics[width=\linewidth]{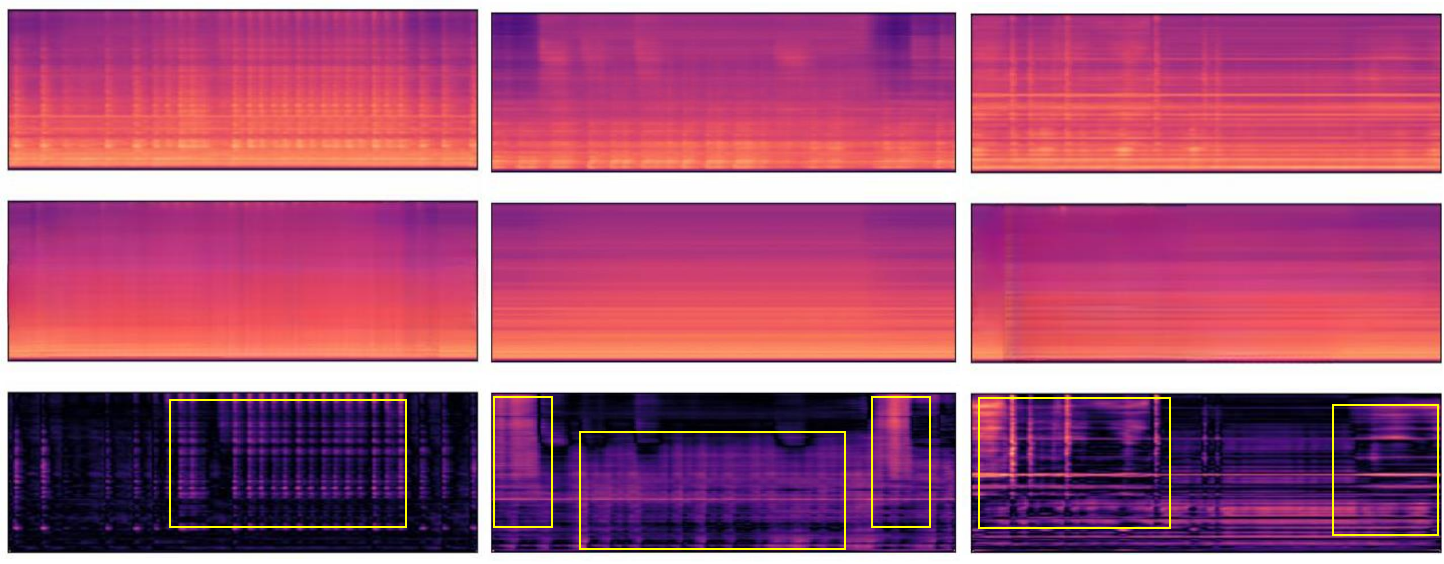}
    {\small (b) ToyCar Anomaly } 
    \label{fig:sub2}
  \end{minipage}
  \vspace{-1mm}
  \caption{Anomaly localization results for normal and anomalous samples of the ToyCar machine in the test set. The first row represents the reconstructed spectrogram from the model, the second row shows the average spectrogram of the training set, and the third row displays the difference between the two. Brighter regions indicate a higher degree of anomaly.}
  \label{fig:3}
    \vspace{-1mm}
\end{figure}

\vspace{-2mm}
Compared to other models, generative reconstruction models demonstrate significant time-frequency localization advantages in industrial acoustic anomaly detection. As shown in Fig. \ref{fig:3}, for normal samples of the ToyCar device (a), the reconstructed spectrogram of the model exhibits high spatial consistency with the average spectrogram of the training set, with only weak, randomly distributed responses in the difference spectrogram. This confirms the model’s precise capability to capture the steady-state operating characteristics of the equipment while maintaining strong robustness against background noise interference. In contrast, for anomalous samples (b), the difference spectrogram displays distinct structured anomaly responses, characterized by both continuous high-intensity band-like patterns and discrete pulse-like bright spots in the time-frequency domain. This effectively demonstrates the model’s ability to disentangle steady-state and transient components within acoustic signals. This further validates the effectiveness of our proposed approach.

\vspace{-5mm}
\section{Conclusion}

\vspace{-2mm}
In this paper, we proposed a novel generative framework, TLDiffGAN, for unsupervised anomalous sound detection in machine sounds. The framework is built upon a Latent Diffusion–GAN (LDGAN) backbone to achieve high-fidelity spectrogram reconstruction, and incorporates a dual-branch structure together with an adaptive temporal mixup module to fuse multimodal acoustic features and enhance sensitivity to subtle temporal patterns. Experimental results on the DCASE 2020 Task 2 dataset demonstrate the superior performance of our approach and its effectiveness in localizing anomalous sound events.

\bibliographystyle{IEEEbib}
\bibliography{paper}

\end{document}